\documentclass[preprint]{emulateapj}
\usepackage{psfig}
\usepackage{apjfonts}
\usepackage{xspace}
\usepackage{amsmath}
\lefthead{Capetti et al.}
\righthead{A VLA survey of early-type galaxies in Virgo}
\begin{document} 

\title{A VERY LARGE ARRAY radio-survey of early-type galaxies in the Virgo cluster}
 
\author{Alessandro Capetti}
\affil{INAF - Osservatorio Astronomico di Torino, Strada
 Osservatorio 20, I-10025 Pino Torinese, Italy}
\author{Preeti Kharb, David J. Axon, David Merritt}
\affil{Department of Physics,
Rochester Institute of Technology, 85 Lomb Memorial Drive,
Rochester, NY 14623, USA}
\and
\author{Ranieri D. Baldi}
\affil{Universit\'a di Torino, via P. Giuria 1, 10125 Torino, Italy}
 
\begin{abstract}
We present the results of a 8.4 GHz Very Large Array radio survey of early-type
 galaxies extracted from the sample selected by C{\^o}t{\'e} and
 collaborators for the Advanced Camera for Survey Virgo Cluster Survey. 
The aim of this survey is to investigate the origin of radio emission in  
early-type galaxies and its link with the host properties in an unexplored
territory toward the lowest levels 
of both radio and optical luminosities. 
Radio images, available for
 all 63 galaxies with B$_{\rm T} < 14.4$, show the presence of a compact radio
 source in 12 objects, with fluxes spanning from 0.13 to 2700 mJy. The
 remaining 51 galaxies, undetected at a flux limit of $\sim$0.1 mJy, have
 radio luminosities L $\lesssim 4 \times 10^{18} \, {\rm W Hz}^{-1}$. The
 fraction of radio-detected galaxies are a strong function of stellar mass,
in agreement with previous results:
 none of the 30 galaxies with M$_{\star} < 1.7 \times 10^{10} M_{\sun}$ is
 detected, while 8 of the 11 most massive galaxies have radio cores. 
There appears to be no simple relation between the presence of a stellar
nucleus and radio emission. In fact, we find radio sources associated with
two nucleated galaxies, but the majority of nucleated objects are not
a radio emitter above our detection threshold.  

A multiwavelength analysis of the active galactic nucleus (AGN) emission,
combining radio and X-ray data, confirms the link between optical surface
brightness profile and radio loudness in the sense that the bright core
galaxies are associated with radio-loud AGNs, while non-core galaxies host
radio-quiet AGNs.  Not all radio-detected galaxies have a X-ray nuclear counter
part (and vice-versa). A complete census of AGNs (and supermassive black holes,
SMBHs) thus requires observations, at least, in both bands. Nonetheless, there
are massive galaxies in the sample, expected to host a large SMBH (M$_{\rm BH}
\sim 10^8 M_{\sun}$), whose nuclear emission eludes detection despite their
proximity and the depth and the spatial resolution of the available
observations. Most likely this is due to an extremely low level of accretion
onto the central SMBH.
\end{abstract}

 
\section{Introduction}
\label{intro}

Nuclear radio emission is almost invariably associated with the presence of an
active galactic nucleus (AGN) and this is an indication that the process of
accretion onto a supermassive black hole (SMBH) naturally produces a signature
in the form of radiation in the radio domain. The separation between
radio-loud (RL) and radio-quiet (RQ) AGNs is in fact only a measure of the
relative flux in the radio band with respect to the optical or X-ray
nucleus \citep{kellermann94,terashima03}; but
also RQ AGNs, when studied at sufficient depth, usually show the
presence at least of a nuclear radio component
\citep[e.g.,][]{ulvestad89,nagar05}. For RL AGNs the radio core results from
synchrotron emission produced by the unresolved base of their jets; for RQ AGNs
the situation is more controversial and it has been recently proposed, besides
the possibility of a jet origin, that their radio nuclei are the manifestation
of the presence of a thermal outflow or of an active disk corona
\citep{blundell07,laor08}

When combined to the very limited effects that absorption has on radio waves,
the study of radio emission provides, in principle, a very powerful tool to
detect accretion onto SMBH and, consequently, 
to establish when a SMBH is present in a
given galaxy. However, to take full advantage of this approach, we must reach
a much deeper understanding of what determines the radio luminosity of a given
galaxy and how this is related to its level of accretion.

A large effort has been dedicated to explore the connection between the host
properties (mostly from an optical point of view) and its radio emission.
Already from the pioneering study by \citet{auriemma77} it was clear that more
massive galaxies have on average a higher radio luminosity than 
smaller galaxies, while apparently there are no distinctions between clusters
and non-clusters members \citep{ledlow96}. The radio luminosity functions
(RLFs) of galaxies of different optical magnitudes are similar but they
differ strongly in their scaling. More recent studies confirm the early
results, indicating that the normalization of the RLF scales with the host
luminosity as $\sim$L$^{2.5}$ \citep[e.g.,][]{best05b,mauch07}. However,
galaxies of given optical magnitude show a very large range of radio power, 
more than 5 orders of magnitude, and
the relation between the radio and optical luminosity can only be described in
terms of a probability distribution.

These studies focused mostly on massive galaxies and had a
relatively high threshold for the radio detection. The analysis reaching the
lowest level of luminosity (in both bands) were performed by \citet{sadler89}
and \citet{wrobel91b},
with limits at L$_{\rm r} \sim 2.5 \times 10^{19} \, {\rm W Hz}^{-1}$ and
M$_{\rm r} \sim $ -18.5. Consequently, the information on the radio
properties of galaxies of lower mass is sparse and incomplete. Similarly,
there is still a dominant fraction of massive galaxies ($\sim 70$ \%)
undetected in radio surveys, for which only an upper limit to their radio
luminosity can be derived.

Clearly, a survey reaching lower radio and optical luminosities can provide
new insights on the origin and properties of the radio emission of early-type
galaxies. The Virgo cluster represents a unique laboratory for such studies.
In fact, it includes hundreds of early-type galaxies, spanning a wide range of
stellar masses, for which a vast suite of data is available in the literature.
In particular, the recent {\it Hubble Space Telescope} ({\it HST}) survey
performed by \citet{cote04} provides us with a detailed analysis of their
optical brightness profiles whose properties, including the presence of a
stellar nucleus, can be included in the study of their radio emission.  Given
its proximity, its members can be observed at high spatial resolution
(1$\arcsec$ at a distance of 17 Mpc corresponds to $\sim$ 80 pc) and their
nuclear emission can be studied down to extremely low luminosity level. For
this reason we performed a radio survey of 63 early-type galaxies in Virgo
with the Very Large Array (VLA). Since the aim of these observations is to explore
their nuclear properties, the data were taken at high resolution (with the
telescope in the A array configuration) and at relatively high frequency (8.4
GHz).

The paper is organized as follows. In Section \ref{sample} we describe the
sample and the VLA observations; in Section \ref{results} we discuss the link
between the radio properties of the galaxies considered with their stellar
mass, with their optical
brightness profile and with the presence of a stellar nucleus; we then perform
an analysis of the multiwavelength properties of their nuclear emission,
taking advantage of X-ray data taken from the literature.
Summary and conclusions are given in Section
\ref{summary}.

\section{Sample selection and VLA observations}
\label{sample}

We considered initially the same sample selected by \citet{cote04} for their
{\it HST} survey of early-type galaxies in the Virgo cluster. More
specifically, they selected early-type galaxies from the VCC catalog
\citep{binggeli87} that consists of 2096 galaxies within this $\approx$ 140
deg$^{2}$ region. A total of 1277 VCC galaxies were considered by
\citet{binggeli87} to be members of Virgo and for 403 cluster members radial
velocities are available. A faint-end cutoff at B$_{\rm T} < 16$ yields 352
galaxies, 163 of which are early-type galaxies according to the VCC
morphological classifications of \citet{binggeli85}. A subset of 100
early-type galaxies was then selected for observations with Advanced Camera
for Survey (ACS) on {\it HST}, excluding objects with uncertain morphologies,
lacking a clearly visible bulge component, with the presence of strong dust
lanes, or signs of strong tidal interactions.

Of the 100 galaxies in the ACS Virgo cluster survey, high-quality radio data
already exist for seven sources, all of them within the top third in terms of
optical luminosity. 
We observed with the VLA additional 56 galaxies,
in order of decreasing B$_{\rm T}$ band magnitude, for a total of 63 objects reaching
B$_{\rm T} = 14.4$. 

\subsection{VLA observations and data reduction}
\begin{figure*}
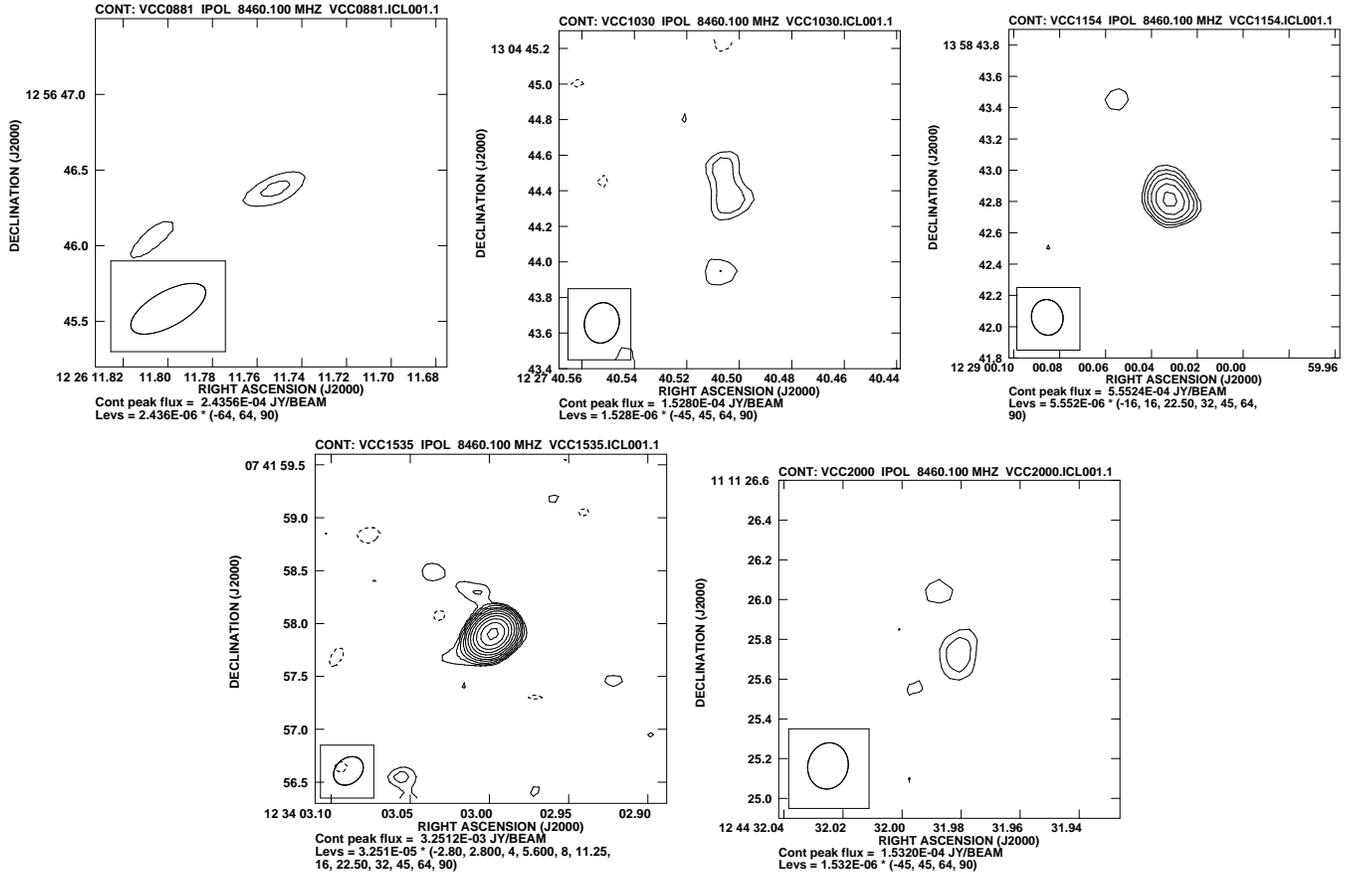

\centerline{
\psfig{figure=VCC0881_new2.ps,width=0.34\linewidth}
\psfig{figure=VCC1030_new2.ps,width=0.33\linewidth}
\psfig{figure=VCC1154_new2.ps,width=0.32\linewidth}}
\centerline{
\psfig{figure=VCC1535_new2.ps,width=0.34\linewidth}
\psfig{figure=VCC2000_new2.ps,width=0.33\linewidth}}
\caption{\label{maps} VLA 8.4 GHz radio maps of the five detected sources. 
The beam size is reported in the bottom left inset. All sources are consistent
with being unresolved, with the exception of
VCC1030, which appears elongated.}
\end{figure*}
The galaxies were observed at 8.4 GHz with the VLA in the A-array 
configuration on 2006 April 17. The source scans, which were
each $\sim$30~minutes long, were interspersed
with $\sim$2~minutes scans of the nearby phase calibrator, 1239+075.
3C286 was used as the flux density calibrator for the experiment.
The data were reduced using the standard calibration and reduction procedures
in the Astronomical Image Processing System (AIPS). 
After the amplitude and phase calibration using the calibrators,
the sources were split from the main data set and imaged using the
task IMAGR. The detected radio sources were weak and were therefore not 
self-calibrated. 

The resolution of the radio images is typically around
$0\farcs3\times0\farcs2$, corresponding to a linear size of
$\sim$ 20 pc at the distance of Virgo.
The typical resulting rms in the radio images is 30 $\mu$Jy beam$^{-1}$.
Five objects were detected with a peak to rms ratio larger than 5,
and radio fluxes of these galaxies 
are reported in Table \ref{tabsample1}, together with the radio core
measurements of the 7 radio galaxies found in the literature.

\begin{table*}
\caption{Properties of the Radio-detected Galaxies}
\label{tabsample1}
%\centering
\begin{tabular}{l c c c c  r r c c c}
\hline\hline
Name & Alt. Name & B$_{\rm T}$ &  $D$ & log M$_\star$ & F$_{\rm core}$ & F$_{\rm 1.4 GHz}$ & Prof. & Nuc.\\
\hline		 		
VCC~1226 & NGC~4472, M~49 & 9.31 & 17.14 & 12.0 & 3.7 $^b$ & 256 & cS & II \\ 
VCC~1316 & NGC~4486, M~87 & 9.58 & 17.22 & 11.8 & 2726$^b$ & 226000 & cS & 0 \\ 
VCC~1978 & NGC~4649, M~60 & 9.81 & 17.30 & 11.7 & 18.4$^c$ & 29 & cS & II \\ 
VCC~0881 & NGC~4406, M~86 &10.06 & 16.83 & 11.9 & 0.25$^a$ & $<$ 1$^d$ & cS & II \\ 
VCC~0763 & NGC~4374, M~84 & 10.26 & 18.45 & 11.7 & 180.7$^b$ & 6100 & cS & II \\ 
VCC~1535 & NGC~4526 & 	10.61 &  16.52 & 11.0 & 3.27$^a$ & 12.00$^d$ & cS$^e$ & 0 \\ 
VCC~1632 & NGC~4552, M~89 & 10.78 & 15.85 & 11.3 & 58.1$^b$ & 100 & cS & II \\ 
VCC~2095 & NGC~4762	 & 11.18  & 16.52 & 10.6 & 1.3$^b$ & $<$ 1$^d$ & S & Ib \\ 
VCC~1154 & NGC~4459 & 	11.37 &  16.07 & 10.9 & 0.57$^a$ & 1.83$^d$ & S & Id \\ 
VCC~1030 & NGC~4435 & 	11.84 &  16.75 & 10.6 & 0.14$^a$ & 2.16$^d$ & S$^e$ & 0 \\ 
VCC~2000 & NGC~4660 & 	11.94 &  15.00 & 10.4 & 0.13$^a$ & $<$ 1$^d$ & S & Id \\ 
VCC~1619 & NGC~4550	 & 12.50  & 15.49 & 10.2 & 0.7$^b$ & $<$ 1$^d$ & S & Ia \\ 
\hline 
\end{tabular} 

\medskip
Notes. Column description: (1) VCC name; (2) alternative optical identification; (3) 
B$_{\rm T}$ magnitude; (4) distance (Mpc); (5)
total stellar mass in solar units; (6) 8.4 GHz radio-core flux (mJy) from
(a) present work, and at 5 GHz from (b) \citet{nagar05}, 
(c) \citet{stanger86}; (7) total 1.4 GHz radio flux (mJy) from 
\citet{condon90,condon02},
or $^d$ from the FIRST survey;
(8) brightness profile classification: cS,
core-S\'ersic; S, S\'ersic; $^e$ classification from our analysis of the
{\it HST}/NICMOS images (see the Appendix); (9) nucleation classes: 
II, non-nucleated; 0, no classification possible; Ia and Ib: nucleated,
Ic, and Id uncertain nuclei.
\end{table*}

\begin{table*}
\caption{The radio undetected galaxies}
\label{tabund}
\begin{tabular}{c c c | c c c| c c c| c c c}
\hline\hline
Name & B$_{\rm T}$ & $F_{\rm core}$$^a$ & Name & B$_{\rm T}$ & F$_{\rm core}$$^a$ & Name & B$_{\rm T}$ & $F_{\rm core}$$^a$
& Name & B$_{\rm T}$ & $F_{\rm core}$$^a$ \\
\hline		 		
VCC0798	& 10.09	&0.11   & VCC1938	& 12.11	&0.10   & VCC1146	& 12.93	&0.08 & VCC1871&	 13.86&	0.09\\
VCC0731	& 10.51	&0.10   & VCC1279	& 12.15	&0.15   & VCC1025	& 13.06	&0.09 & VCC0009&	 13.93&	0.08\\
VCC1903	& 10.76	&0.10   & VCC1720	& 12.29	&0.11   & VCC1303	& 13.10	&0.09 & VCC0575&	 14.14&	0.08\\
VCC1231	& 11.10	&0.13   & VCC0355	& 12.41	&0.10   & VCC1913	& 13.22	&0.09 & VCC1910&	 14.17&	0.12\\
VCC1062	& 11.40	&0.08   & VCC1883	& 12.57	&0.10   & VCC1327	& 13.26	&0.17 & VCC1049&	 14.20&	0.17\\
VCC2092	& 11.51	&0.08   & VCC1242	& 12.60	&0.09   & VCC1125	& 13.30	&0.10 & VCC0856&	 14.25&	0.10\\
VCC0369	& 11.80	&0.08   & VCC0784	& 12.67	&0.08   & VCC1475	& 13.36	&0.26 & VCC0140&	 14.30&	0.08\\
VCC0759	& 11.80	&0.07   & VCC1537	& 12.70	&0.08   & VCC1178	& 13.37	&0.14 & VCC1355&	 14.31&	0.08\\
VCC1692	& 11.82	&0.07   & VCC0778	& 12.72	&0.08   & VCC1283	& 13.45	&0.13 & VCC1087&	 14.31&	0.08\\
VCC0685	& 11.99	&0.09   & VCC1321	& 12.84	&0.08   & VCC1261	& 13.56	&0.12 & VCC1297&	 14.33&	0.15\\
VCC1664	& 12.02	&0.09   & VCC0828	& 12.84	&0.08   & VCC0698	& 13.60	&0.11 & VCC1861&	 14.37&	0.10\\
VCC0654	& 12.03	&0.11   & VCC1250	& 12.91	&0.08   & VCC1422	& 13.64	&0.10 & VCC0543&	 14.39&	0.10\\
VCC0944	& 12.08	&0.10   & VCC1630	& 12.91	&0.08   & VCC2048	& 13.81	&0.09 &        &        &   \\
\hline 
\end{tabular} 

\medskip
$^a$ 3 $\sigma$ upper limits in mJy.
\end{table*}

Radio images of these five sources are reproduced in Figure \ref{maps}.  The
radio structures are typically unresolved and core-like; the exception is
VCC1030, which appears elongated over $\sim 0\farcs2$ approximately along the
direction of its large-scale dusty disk. The position of the radio source is
always within less than $0\farcs$8 from the galaxy's center as measured in the
{\it HST} images, and consistent with being coincident with it considering the
uncertainties in the relative astrometry.

None of the observed galaxies is detected in the FIRST survey (with a
  limiting point source flux density of $\sim$ 1 mJy) with the exceptions of
  VCC~1535, VCC~1154, VCC~1030, that are also seen in our images. 
For the
  undetected sources we provide the corresponding upper limits in Table
  \ref{tabund}.

\begin{figure*}
\centerline{
\psfig{figure=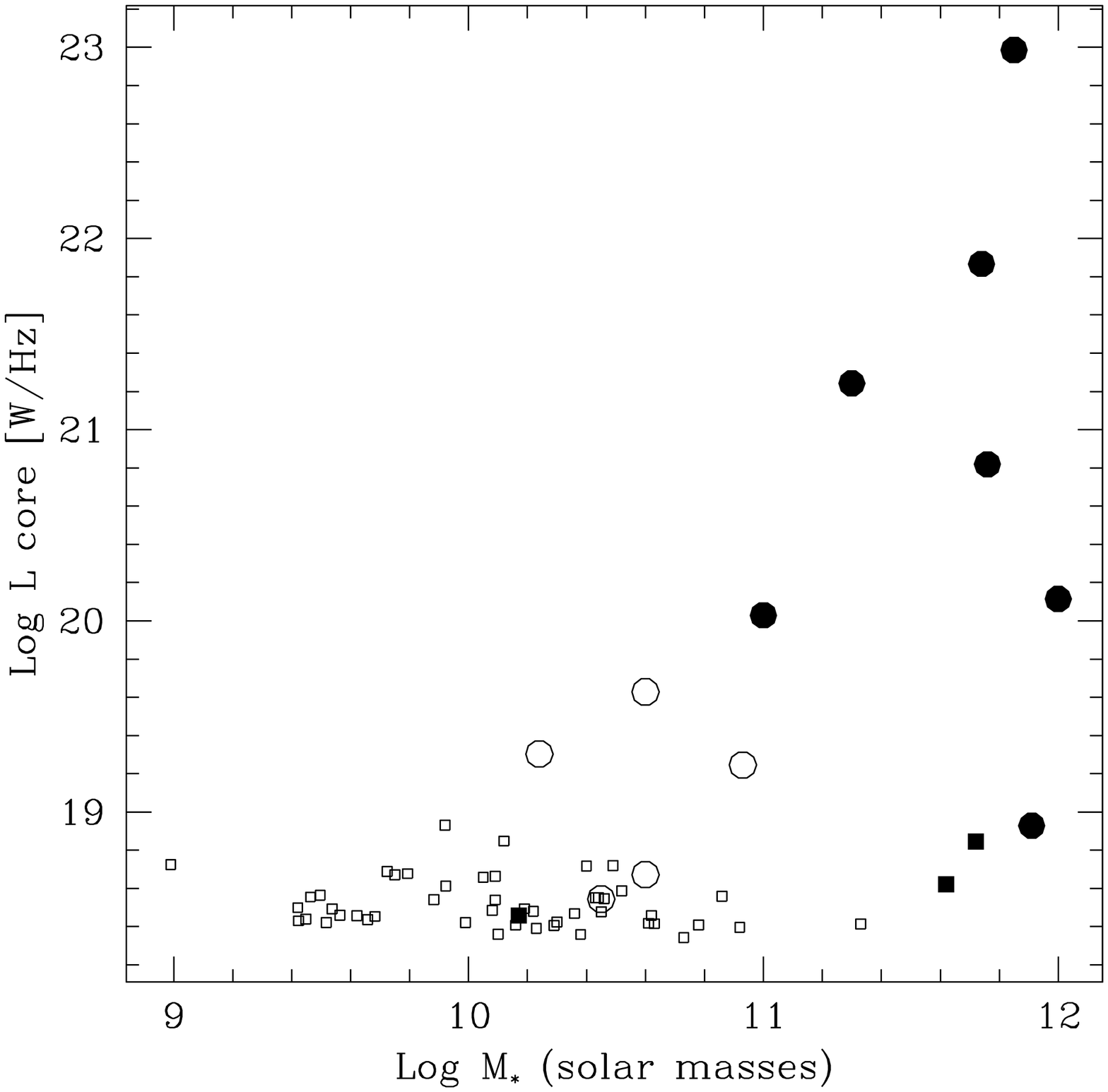,width=0.50\linewidth}
\psfig{figure=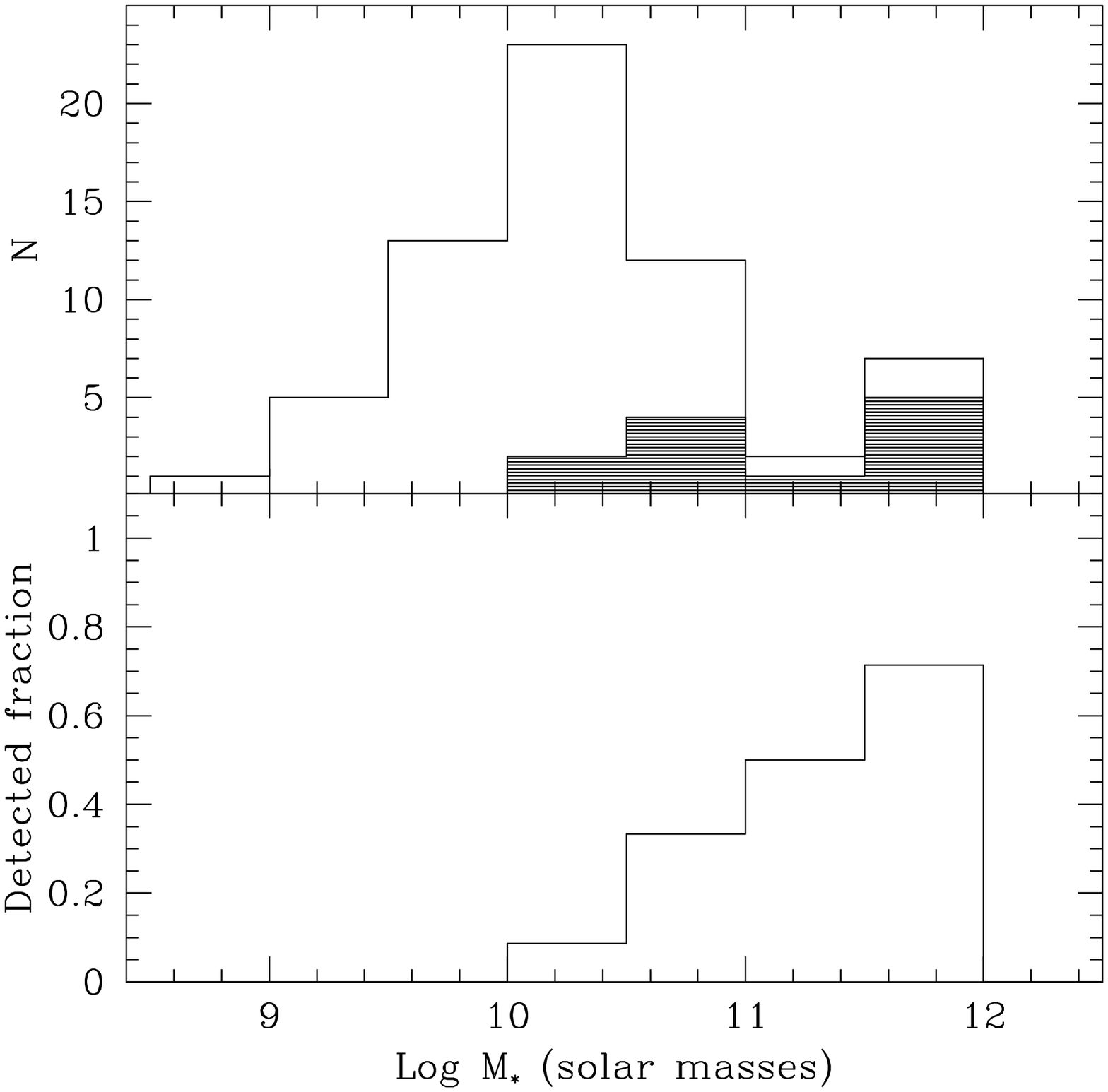,width=0.50\linewidth}}
\caption{\label{firr} Left: radio core luminosity vs stellar masses. The radio
 detected galaxies are marked with large circles, separating them on the
 basis of the optical surface brightness profiles: filled circles are
 core galaxies, while empty circles are pure S\'ersic
 galaxies. The undetected objects are marked with small squares (for
 clarity, we omit the downward pointing arrows), empty for S\'ersic galaxies,
 filled for core galaxies.  
 Right-top panel: distribution of stellar masses for the whole sample (empty
 histogram) and for the detected sources (filled histogram); bottom panel:
 fraction of detected sources vs. stellar mass in 0.5 dex bins.}
\end{figure*}

\section{Results}
\label{results}

\subsection{Radio emission and host stellar mass}
\label{bivariate}

In Figure \ref{firr}, left panel, we show the 3.6 cm radio luminosity of the VCC
sources against their stellar
masses, estimated using the recipe of \citet{bell07}
$${\rm log} \,(M_\star/L_{g_0}) = 0.698 (g_0 - z_0) - 0.367$$
where $g_0$ and $z_0$ are the extinction corrected Sloan Digital Sky Survey (SDSS) total magnitudes.
We adopted the distances estimated by \citet{mei07} from the analysis of the
surface brightness fluctuations (see Table \ref{tabsample1}) 
and set the radio upper limits at 3 times the rms noise in the images.
The typical radio flux limit of $\sim$ 1 mJy corresponds to a radio
luminosity of L $\sim 4 \times 10^{18} \, {\rm W Hz}^{-1}$. 

The radio detections are concentrated at the high end of the mass
distribution, with no detections for the 30 galaxies with M$_{\star} < 1.7
\times 10^{10}M_{\sun}$ (or $M_{\rm B} > -18.6$). The fraction of detected
galaxies increases with increasing stellar mass (Figure \ref{firr}, right
panel), reaching $\sim$70 \%. This is qualitative agreement with previous
results on the bivariate radio-optical luminosity function of early-type
galaxies that show that the fraction $f$ of galaxies brighter than a given
radio luminosity $L$ is a strong function of the host mass, i.e. $ f \propto
{\rm M}_{\star}^{2.0 - 2.5}$ \citep[e.g.,][]{best05b,mauch07}.

More quantitatively, we estimated the number of objects of our sample expected
to be detected, assuming that the \citet{best05b} probability law, derived
from SDSS selected galaxies, can be extended to our regime of lower masses and
radio-luminosities.  We computed the probability of a radio detection for each
source and derived the number of expected detections up to a given host mass.
At the low mass end this predicts the detection of 0.06 sources up to M$_\star
< 1.7 \times 10^{10}M_{\sun}$, in agreement with the lack of radio
detections in the 30 fainter galaxies of the sample. Considering now the high
end mass, the probability law saturates to 100\% at M$_\star \sim
10^{11}M_{\sun}$, implying that all galaxies above this level are expected
to be seen at the depth of our radio images. This is not the case since we
have three radio undetected, optically bright, galaxies.  Previous studies,
characterized by far higher radio luminosity thresholds concur that a
saturation of the detection rate at a $\sim$ 30 \% level occurs at the high
mass end \citep[e.g.,][]{mauch07}. Not surprisingly, expanding the radio
luminosity coverage downward by 4 orders of magnitude we reach a higher
detection rate, $\sim$ 70 \%. Nonetheless, we still have three bright galaxies
without a radio source associated with them. These objects are $\gtrsim 30000$
times fainter than the radio core of M~87 (the brightest VCC galaxy)
and at least $\gtrsim 200000$
times fainter in terms of total radio power, despite the similar stellar mass.
This result emphasizes the probabilistic nature of the radio-optical bivariate
law and that the optical luminosity of a given galaxy is not a good predictor
of its level of radio emission.

\subsection{Nucleation and radio emission}
\label{nucleation}

From the analysis of the data-set of {\it HST} images, \citet{cote06} found that a
large fraction of VCC early-type galaxies is nucleated, with a frequency of 66 \%
- 82 \%. Core galaxies do not follow this general rule, since they lack
resolved stellar nuclei but, conversely, they often show unresolved optical
nuclear sources (defined as nucleation class II). The origin of the nuclei in
this class of galaxies, many of them associated with bright radio-sources, must
be ascribed to the active nucleus and most likely they represent the
synchrotron emission from the basis of their radio jets
\citep{chiaberge:ccc,balmaverde06b,cccpol}.

We examine the possible presence of a link between nucleation and radio
emission. Leaving aside the core galaxies, we are left with only five galaxies
with a radio detection and a S\'ersic profile. There are two clear nuclei
(classes Ia and Ib), two uncertain nuclei (class Ic and Id), and a dusty galaxy (VCC1030)
that cannot be classified from the point of view of nucleation (class 0). In
the whole sample of 100 VCC the breakdown in terms of nucleation is 62:15:12:6
in the classes Ia-b:Ic-d:II:0 respectively\footnote{With an additional five
 sources of class Ie, reserved for compact sources offset from the galaxy
 center.}.

The stronger connection between nucleation and radio properties is therefore
the high incidence of radio nuclei in the core galaxies of the sample, lacking
stellar nuclei. Leaving aside the core galaxies, the statistics is clearly
very poor, but apparently there is no simple relation between nucleation and
the detection of a radio source. In fact, we find radio sources associated with
two certainly nucleated galaxies, but the majority of nucleated objects are not
radio emitters above our detection threshold.  Furthermore, the two nucleated
galaxies for which we have a radio detection are among the brightest sources of
our sample.

These results are consistent with the study by \citet{seth08} on the
coincidence of nuclear star clusters and activity.  They found that nucleated
galaxies can host an AGN (thus a star cluster and an AGN are not mutually
exclusive) and the fraction of nucleated active increases strongly with
increasing galaxy mass.
However, since this result
applies also to the general population of galaxies, regardless of their
nucleation, they concluded that the presence of a stellar nucleus is not
linked to nuclear activity.

%\subsection{Radio emission and properties of the host brightness profile}
%\label{sbp}

\subsection{Host brightness profile and nuclear multiwavelength properties}
\label{chandra}

\begin{table}
\caption{Comparison of X-ray and radio nuclear luminosities
 of VCC galaxies with both
 {\it Chandra} and VLA observations available.}
\label{lx}
\centering
\begin{tabular}{l c c c}
\hline\hline
Name & L$_x$ (0.3 - 10 keV) & $\nu L_r$ & Profile \\
\hline	 	 
VCC~1226& $<$38.49 & 36.81 & cS\\ 
VCC~1316& 41.20 & 39.68 & cS\\ 
VCC~1978& 39.05 & 37.52 & cS\\ 
VCC~763 & 39.73 & 38.57 & cS\\ 
VCC~1632& 39.58 & 37.94 & cS\\ 
VCC~2095& 38.71 & 36.33 & S \\ 
VCC~0881& $<$38.64 & 35.85 & cS\\ 
VCC~1535& $<$38.21 & 36.95 & cS \\ 
VCC~1154& 39.03 & 36.17 & S \\ 
\hline
VCC~798 & $<$38.43	& $<$35.55 & cS \\
VCC~731 & 39.00$^a$ & $<$35.54 & cS \\
VCC~1903 & 39.11	& $<$35.34 & S \\
VCC~1231 & 38.60	& $<$35.48 & S \\
VCC~2092 & 38.59	& $<$35.32 & S \\
VCC~1692 & 38.45	& $<$35.34 & S \\
VCC~685 & 39.14	& $<$35.38 & S \\
VCC~1664 & 39.95	& $<$35.34 & S \\
VCC~1720 & $<$38.54	& $<$35.48 & S \\
VCC~1883 & 38.35	& $<$35.47 & S \\
VCC~1913 & $<$38.46	& $<$35.46 & S \\
VCC~1178 & 38.67	& $<$35.54 & S \\
VCC~2048 & $<$38.12	& $<$35.38 & S \\
VCC~9 & $<$38.15	& $<$35.38 & S \\
VCC~1049 & $<$38.08	& $<$35.65 & S \\
VCC~1297 & 38.42	& $<$35.61 & S \\
\hline
\end{tabular}

\medskip
$^a$ The identification of the brightest X-ray source with the AGN is uncertain
\citep{sivakoff03}.
\end{table}

The ACS/{\it HST} Virgo survey images were used by \citet{ferrarese06} to explore
the properties of the surface brightness profiles (SBP) of the VCC galaxies.
They found that, while the SBP are in general well described by a
\citet{sersic68} model, in most of the brightest galaxies the inner profiles
are lower than expected based on an extrapolation of the outer S\'ersic law.
These galaxies are better described by a core-S\'ersic profile as defined by
\citet{trujillo04}\footnote{Also the low luminosity VCC~1250 is classified as
  a core-galaxy but this classification is considered as marginal and should
  be interpreted with caution since the galaxy morphology is severely affected
  by dust patches and a bright nuclear source.}.
Table \ref{tabsample1} reports
the complete list of optical classifications.\footnote{
In the Appendix we derive a classification of two galaxies of our sub-sample
(VCC~1030 and VCC~1535) from new infrared {\it HST}. Their optical images are
seriously contaminated by the presence of a large-scale dusty disk that prevent
the study of the SBP.}

%For 7 out of 9 of the core galaxies in our sample
%we have a radio detection, while VCC~731 and VCC~798 remain undetected. 
%Conversely only 5 out 54 S\'ersic galaxies are detected in the VLA images but
%they do not exceed a radio core luminosity of $4 \times 10^{19}$ W/Hz
%and a total radio power at 1.4 GHz of $2 \times 10^{20}$ W/Hz.

%\subsection{Host brightness profile and nuclear multiwavelength properties}
%\label{chandra}

\begin{figure}
\centerline{
\psfig{figure=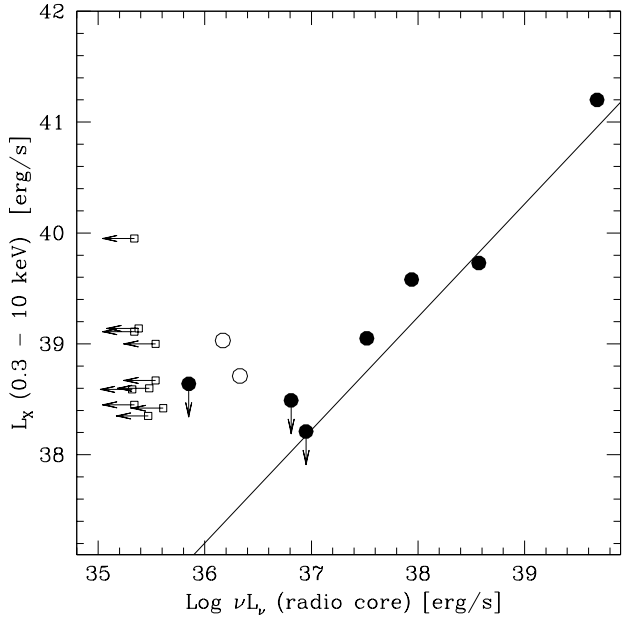,width=1.00\linewidth}}
\caption{\label{lxlr} Comparison of radio and X-ray nuclear luminosities from
 \citet{gallo08}. Filled circles are core galaxies and empty circles are
 S\'ersic galaxies. The empty squares
 correspond to X-ray nuclei in galaxies not detected in our VLA
 survey. For clarity we do not show the objects that are 
upper limits both in radio and X-ray. The
 solid line represents the correlation of radio and X-ray luminosity for
 RL AGNs found by \citet{balmaverde06b}.}
\end{figure}

\citet{gallo08} presented preliminary results of a program of 
{\it Chandra} observations of the VCC sample.
They report the analysis of X-ray images for 32 objects, 25 in common with our
sub-sample, and found X-ray nuclei in 16 galaxies,
whose luminosities are given
in Table \ref{lx}. They conclude that these
nuclear sources are most likely the manifestation of the presence of a
low luminosity active nucleus. 
This opens the possibility of exploring the link between
host brightness profile and AGNs multiwavelength properties.

{\it Chandra} observations revealed an X-ray nucleus in five core galaxies, and all
of them\footnote{With the only apparent exception of VCC~731.  However,
  \citet{sivakoff03} stated that the identification of the brightest X-ray
  source with the AGN is uncertain.} are consistent with the correlation
linking radio and X-ray emission for RL AGNs derived by
\citet{balmaverde06b} (see Figure \ref{lxlr}). Moving to the S\'ersic galaxies,
two have nuclei detected both in radio and X-ray, while
10 are only detected in the X-ray images. The location of these sources in the
L$_{\rm r}$ versus L${\rm_X}$ plane indicates that they are low luminosity
RQ AGNs. These results confirm the relation between optical surface
brightness profile and radio loudness \citep{paper3,seyfert07} in the sense
that core galaxies are associated with RL AGNs, while non-core galaxies
host RQ AGNs.

A consequence of these findings is that, not surprisingly, it is easier to
detect the AGN emission in the radio band for a RL source and in the X-ray
band for a RQ AGNs. This has an important consequence when we seek for a
complete census of AGNs and consequently of SMBHs in a sample of galaxies,
since only by combining radio and X-ray data it is possible to cover all AGNs
manifestations. For example, the presence of an AGN in VCC~1226, the brightest
cluster member, is not visible in X-ray images, while it is clearly shown by
its radio images.

\section{Summary and conclusions}
\label{summary}

We presented the results of a radio survey of early-type galaxies in the Virgo
cluster, extracted from the sample selected by \citet{cote04} for their ACS
Virgo Cluster Survey. We observed 56 galaxies at 8.4 GHz with the VLA that,
combined with data from the literature, provide radio images for all 63
galaxies brighter than B$_{\rm T} = 14.4$. 
The aim of this survey is to investigate the origin of radio emission in  
early-type galaxies and its link with the host properties in an unexplored
territory toward the lowest levels of both radio and optical luminosities. 

Compact radio sources are found in 12 objects, with fluxes from 0.13 to 2700
mJy. The remaining 51 galaxies are undetected at a flux limit of $\sim$0.1
mJy, corresponding to radio luminosities L $\lesssim 4 \times 10^{18} \, {\rm
  W Hz}^{-1}$. The fraction of radio-detected galaxies is a strong function
of stellar mass, in agreement with previous results on the bivariate
radio/optical luminosity function of early-type galaxies: while none of the 30
galaxies with M$_{\star} < 1.7 \times 10^{10} M_{\sun}$ is detected, 8 of
the 11 most massive galaxies have radio cores.  However, the galaxy mass is
not a good predictor of its level of radio emission. This is clearly seen from
a comparison of the radio properties of the brightest galaxies of the VCC.
While they span only a factor of 10 in stellar mass, they cover a range of
$\gtrsim$ 4.5 orders of magnitude in radio-core power, and three of them are
not detected in our radio images.

We note that VCC early-type galaxies with strong dust lanes or of
  signs of interactions were excluded by the original definition of the sample
  for the ACS Virgo cluster survey. Considering the links between mergers,
  dust content and nuclear activity suggested by previous studies of
  early-type galaxies \citep[e.g.,][]{tran01,deruiter02,lauer05} the sub-sample
  considered might be biased against active galaxies with respect to the
  overall population. This effect, however, does not affect the bright
  luminosity end, since the sample is complete down to B$_{\rm T}$ = 12.

We examined the possibility of a link between nucleation and radio
emission. Core galaxies lack stellar nuclei and show a high incidence
of radio nuclei. In the rest of the sample, only two nucleated
galaxies show the presence of a radio source. Thus nuclear activity
can coexist with a stellar nucleus, but the fraction of active
galaxies is not related to nucleation.

We considered the properties of the surface brightness profiles, separating
galaxies reproduced by a S\'ersic law from those better described by a
core-S\'ersic profile. Furthermore, we relied on the results of X-ray {\it Chandra}
observations by \citet{gallo08} to derive a multiwavelength view of the AGN
emission. The reported link between optical surface brightness profile and
radio loudness \citep{paper3} is found also in the VCC sample, since
RL AGNs are associated with core galaxies, while non-core galaxies only
host RQ AGNs.  

Not all radio-detected galaxies have a X-ray nuclear counter part, and
vice-versa. Not surprisingly, it is easier to detect the AGN emission in the
radio band for a RL source and in the X-ray band for a RQ
AGNs. This has an important consequence when we seek for a complete census of
AGNs and consequently of super-massive black holes for which a combination of
(at least) radio and X-ray data is required.

Nonetheless, nuclear emission is not detected in a significant fraction of VCC
galaxies in either observing band. Clearly, there is the possibility that the
VCC sample crosses the minimum galaxy mass at which a SMBH
is present (if such threshold indeed exists). However, there are no signs for
the presence of an AGN in relatively massive VCC galaxies (e.g., VCC~798 and
VCC~1720) despite the fact that they are expected to host a large SMBH
(M$_{\rm BH} \sim 10^8M_{\sun}$), based on a typical ratio of 0.002 between
SMBH and galaxy's mass \citep{marconi03}.
This is particularly worrisome for a
general use of AGNs as black-hole tracers, considering the proximity of the
Virgo cluster, the depth and the spatial resolution of the available
observations. 

In general, we cannot distinguish for the quiescent galaxies between a
scenario 1) of a transition toward
galaxies lacking a SMBH due to a low mass threshold, 2) the possibility
that the SMBH has been ejected due to the recoil caused by the coalescence
between two black holes following a galaxies merger
\citep[e.g.,][]{campanelli07}, and 3) of extremely low accretion rate.

In fact, a crucial role in our ability to detect an active nucleus, and in
setting its level of activity, is certainly played by the level of accretion
onto the central SMBH. At least for low-power RL AGNs, it has 
shown that a suggestive linear correlation exists between the
accretion rate of hot gas (estimated in a spherical approximation) and the jet
power, over several orders of magnitude \citep{allen06,balmaverde08}.
Therefore, to take full advantage of active nuclei as black hole tracers, it
is necessary to combine a multiwavelength approach with estimates of the
accretion rate. It is certainly of great interest to explore for example
whether the connection between the level of accretion and activity can be
extended to the lowest level of radio emission seen in the RL VCC
galaxies, and how this compares with similar estimates for RQ AGNs.

\acknowledgements We thank the anonymous referee for his/her comments that
help to improve the paper.  The National Radio Astronomy Observatory is a
facility of the National Science Foundation operated under cooperative
agreement by Associated Universities, Inc. Based on observations obtained at
the Space Telescope Science Institute, which is operated by the Association of
Universities for Research in Astronomy, Incorporated, under NASA contract NAS
5-26555.  This research has made use of the NASA/ IPAC Infrared Science
Archive, which is operated by the Jet Propulsion Laboratory, California
Institute of Technology, under contract with the National Aeronautics and
Space Administration.

\appendix

The optical images of two galaxies of the sample (namely VCC~1030 and
VCC~1535) both detected in the VLA images, show the presence of large-scale
dusty disks \citep{cote04} that prevent the study and classification of their
surface brightness profiles. We then retrieved from the {\it HST} archive their
infrared images where the impact of dust absorption is less severe. The
images were taken with NICMOS/{\it HST} through the filter F160W ({\it H} band)
and were processed by the standard {\it HST}
pipeline. The camera NIC1 was used, with a pixel size of 0\farcs043, for
a field of view of $\sim$ 11\arcsec $\times$ 11\arcsec. 

\begin{figure*}
\centerline{
\psfig{figure=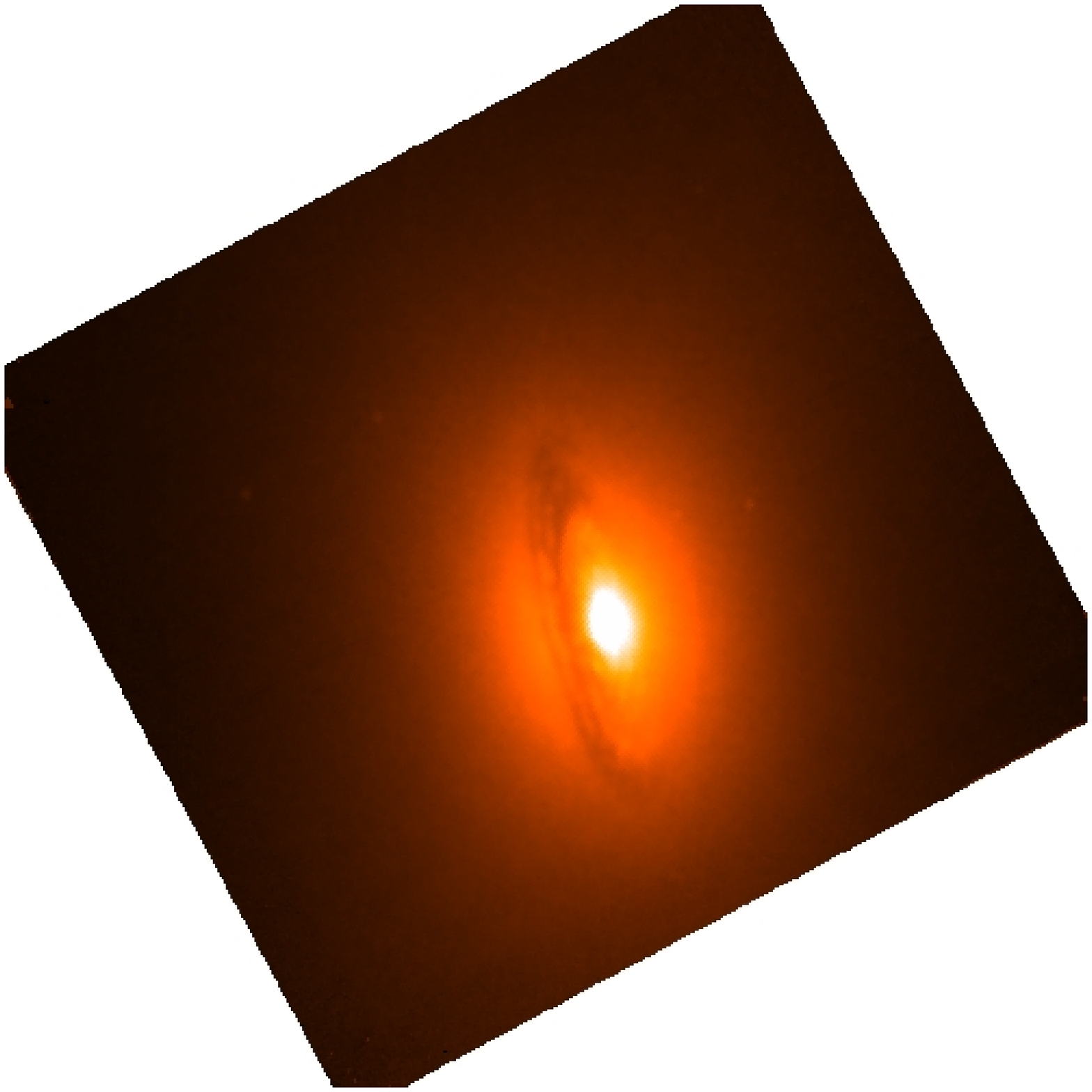,width=0.40\linewidth}
\psfig{figure=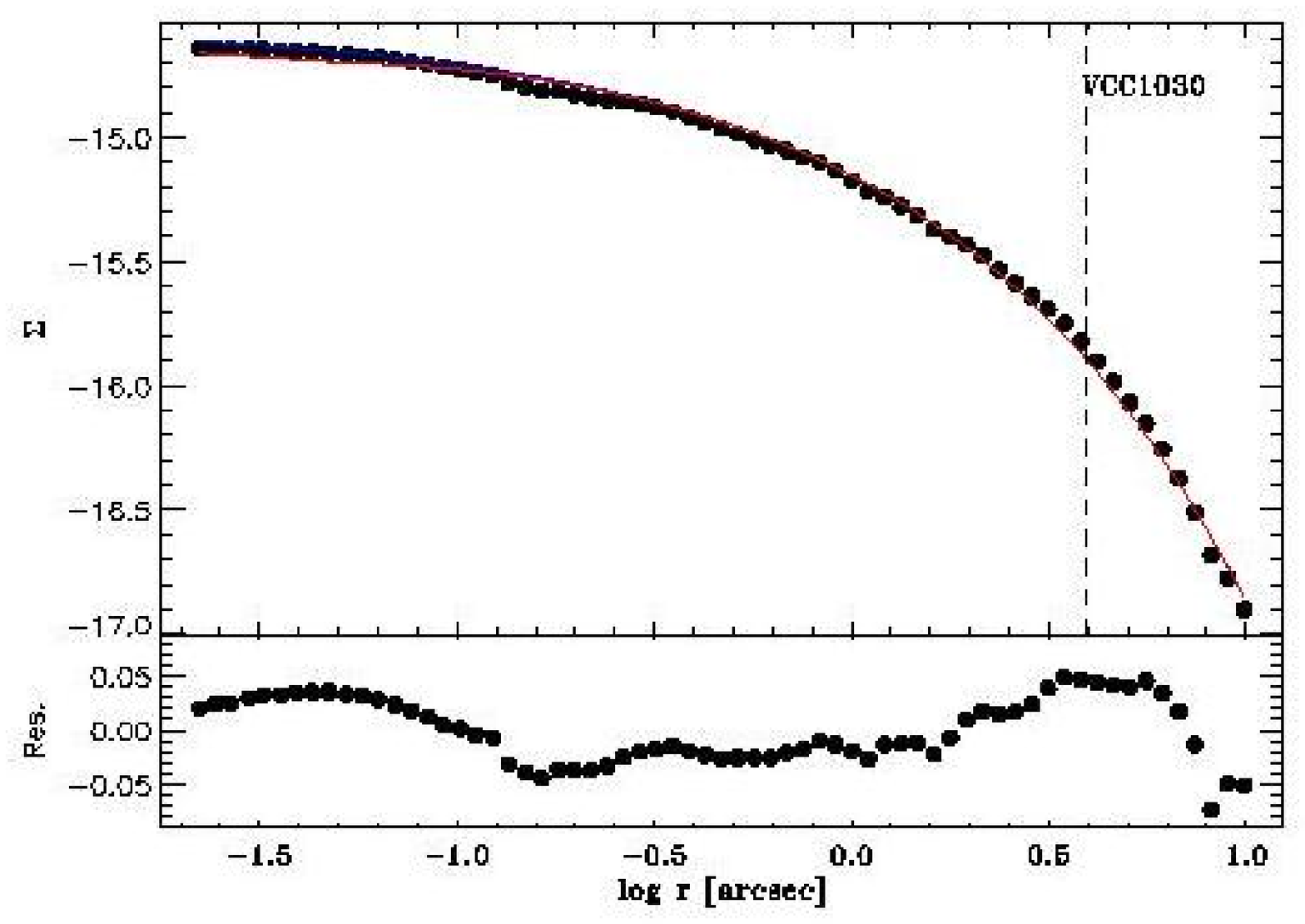,width=0.50\linewidth}}
\caption{\label{sbpfig1} {\it HST}/NICMOS image of VCC~1030 
and (on the right panel) the derived surface
brightness profile. The solid line reproduces the best fit S\'ersic law.
The vertical dashed
line marks the value of the effective radius, $r_e = 3.9\arcsec$.
The vertical axis gives the logarithm of the surface brightness in unit of
[erg s$^{-1}$ cm$^{-2}$ \AA$^{-1}$ arcsec$^{-2}$].
In the bottom inset we show the residuals of the fit.}
\end{figure*}

\begin{figure*}
\centerline{
\psfig{figure=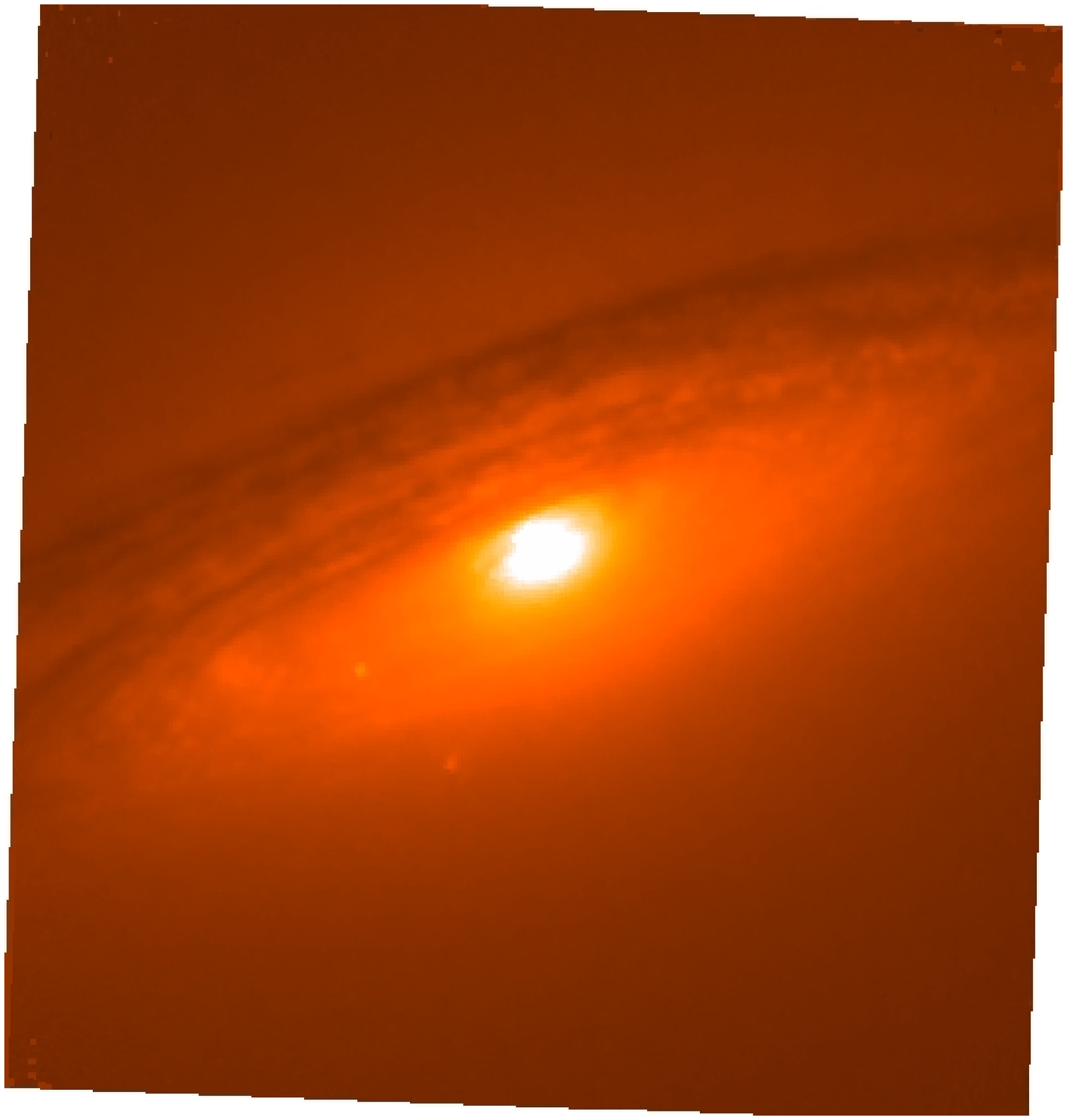,width=0.35\linewidth}
\psfig{figure=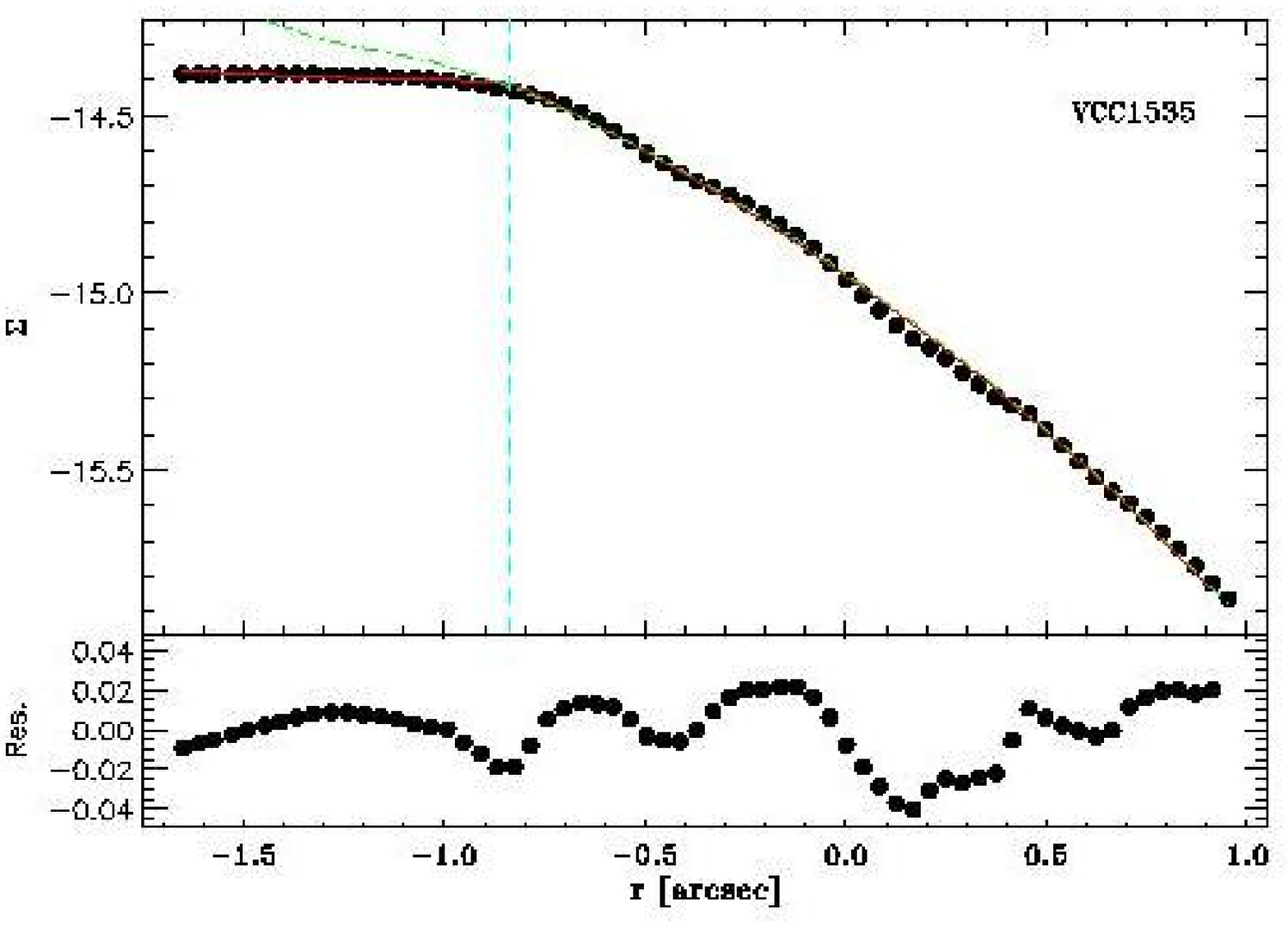,width=0.50\linewidth}}
\caption{\label{sbpfig2} Same as Figure \ref{sbpfig1} for VCC~1535.  The
  vertical dashed line marks the value of the core radius, $r_c =
  0\farcs15$. The dot-dashed line is the best-fit S\'ersic law to the external
  regions ($r > 1\arcsec$) of the galaxy, that overpredicts the central
  surface brightness.}
\end{figure*}

Elliptical isophotes were fit to both images using the IRAF task `ellipse'
\citep{jedrzejewski87}. Although these images are still affected by dust
absorption (see Figure \ref{sbpfig1} and \ref{sbpfig2}), there are sufficient
dust free regions to derive the SBP after proper masking. Since the image of
VCC~1535 fills completely the field of view, we extended the radial coverage
of the SBP using the {\it H} band image from the Two Micron All Sky Survey
(2MASS).

The SBP were fit using a S\'ersic law \citep{sersic68} convolved with the
appropriate point-spread function before comparison with the data. The SBP of
VCC~1030 is reproduced very closely by a S\'ersic model with an effective
radius of $r_e = 3\farcs9$ and a S\'ersic index $n=1.7$. This is not the
case of VCC~1535 that shows a strong light deficit in the innermost regions
with respect to the S\'ersic that describes the external regions and requires
the presence of a flat central core. We therefore fit its SBP with a
core-S\'ersic model \citep{trujillo04}; the parameters of the best fit are an
effective radius $r_e = 106\arcsec$, an index $n= 5.2$, a core radius $r_c =
0\farcs15$, and a inner slope $\gamma = 0.06$. This analysis leads to a
classification based on the SBP properties of VCC~1030 as a S\'ersic galaxy
and of VCC~1535 as a core-S\'ersic galaxy.

\end{document}